\documentclass[runningheads]{llncs}
\usepackage{graphicx}
\usepackage{hyperref}   
\usepackage{url}    
\hypersetup{
    colorlinks=true,
    linkcolor=blue,
    filecolor=magenta,      
    urlcolor=magenta,
}
 
\begin{document}
\title{Regression and Learning with Pixel-wise Attention for Retinal Fundus Glaucoma Segmentation and Detection} %\thanks{Supported by organization x.}}
\author{Peng Liu\inst{1} \and
% Yaxin Shen\inst{2} \and
% Ling Dai\inst{2} \and
% Yan Chen\inst{3} \and
% Weiping Jia\inst{3} \and
% Huating Li\inst{3} \and
% Bin Sheng\inst{2} \and
Ruogu Fang\inst{1}}
\authorrunning{L. Author et al.}
\institute{J. Crayton Pruitt Family Dept. of Biomedical Engineering
University of Florida, USA}
\maketitle              % typeset the header of the contribution
\begin{abstract}
% Glaucoma is a common type of eye disease and currently the leading factor of irreversible vision loss in the world. 
Observing retinal fundus images by an ophthalmologist is a major diagnoses approach for glaucoma.  However, it is still difficult to distinguish the lesions features solely through manual observations, especially, in glaucoma early phase.  In this paper, we present two deep learning based automated algorithms for glaucoma detection and optic disc and cup segmentation. We utilize the attention mechanism to learn pixel-wise features for accurate prediction.  In particular,  we present two convolutional neural networks that can focus on learning various pixel-wise level features.  In addition, we develop several attention strategies to guide the networks to learn the important features that have a major impact on prediction accuracy.   We evaluate our methods on validation dataset and The proposed both tasks' solutions can achieve  impressive results and outperform current state-of-the-art methods.  \textit{The code is available at \url{https://github.com/cswin/RLPA}}.  

\end{abstract}
\section{Introduction}
This work is a workshop challenge- Retinal Fundus Glaucoma Challenge (REFUGE). The goal of the challenge is to evaluate and compare automated algorithms for glaucoma detection and optic disc/cup segmentation on a common dataset of retinal fundus images. We proposed two solutions that achieved the top performance for both segmentation and classification tasks. 
The solutions have the potential to be extended to either a novel methodology or an application.  The details can be found from https://refuge.grand-challenge.org/Home/ or the paper~\cite{orlando2020refuge} published on TMI2019.
\section{Method}

\subsection{Segmentation}

\begin{figure}\label{xunet}
\includegraphics[width=\textwidth]{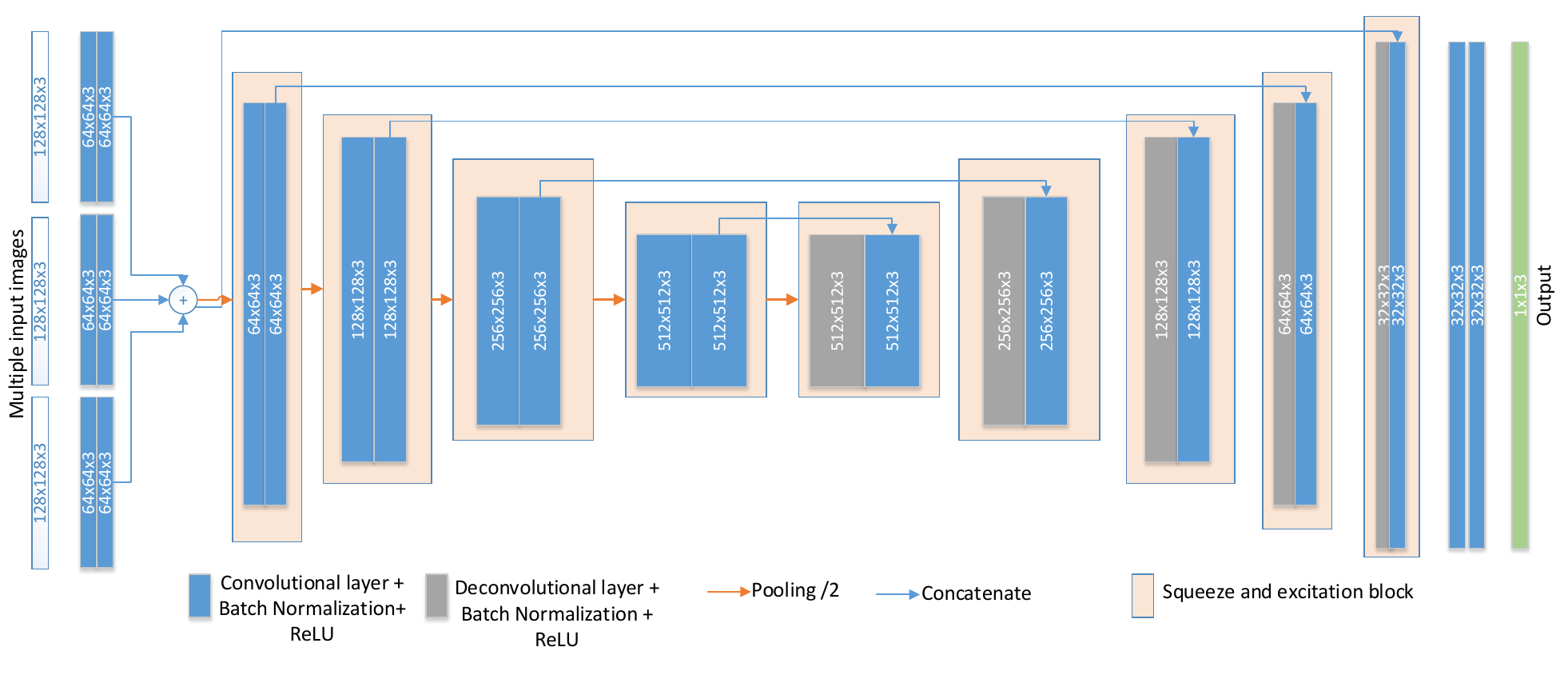}
\caption{X-Unet's architecture includes squeeze-and-excitation blocks.} 
\end{figure}

We employ a u-net~\cite{ronneberger2015u} like architecture to learn the different pixel-level features.  We modify the u-net to have multiple inputs (3 in our case) so that the network can receive more original raw pixel information during training.  This strategy can reduce the risk of overfitting and enhance the network's learning capability.  We refer to this architecture as X-Unet.    Moreover, we embed the squeeze-and-excitation blocks~\cite{hu2017squeeze} into our X-Unet to weight the features from different convolutional layers' channels.  In particular, 
 we utilize a mechanism that allows the network to selectively amplify the valuable channel-wise features and suppress the useless feature from global information. In addition, we use deconvolution in the network decoder part to refine the decoding capability by refusing the features between different level encoded features and the corresponding level decoded features.  The figure~\ref{xunet} shows our X-Unet's architecture. 

\subsubsection{Regression}
We consider the segmentation task as an image regression instead of pixel classification problem, which in deep learning usually needs to transform the low-level pixel information to high-level features. However, for the disc and cup binary segmentation tasks, low-level pixel-wise features are more important. In contrast to learning to classify the pixels, mapping a retinal image to its corresponding label directly can keep more low-level pixel-wise features.

\begin{figure}\label{regression}
\includegraphics[width=\textwidth]{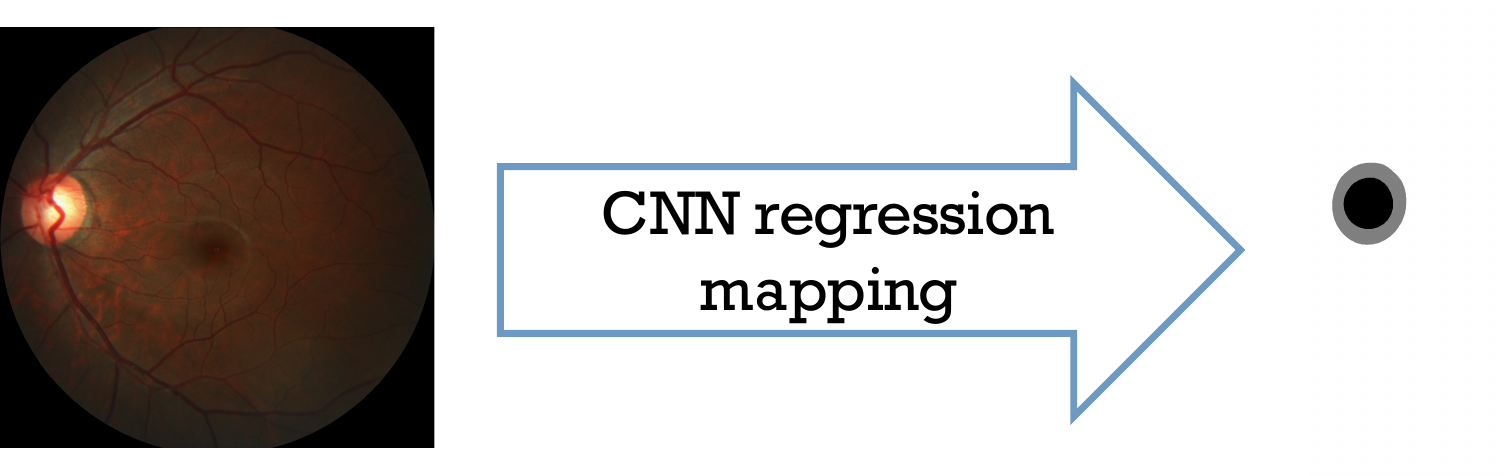}
\caption{Consider the segmentation as an image regression instead of pixel-classification problem.} 
\end{figure}

\subsubsection{Loss function}

The major pixel-wise similarities in training images allow us to adopt mean absolute error (MAE) as our loss function to calculate the pixel-wise difference between label and prediction. 
% In addition,  unlike existing work directly adding weights into each pixel-class loss calculation to release the unbalanced pixel samples (e.g., background v.s optic disc/cup)  in training images, we first develop a Split-Copy-Merge strategy to enable multiple  X-Unets to predict different pixel-wise class mask separately, and then add two weights for mask pixels ($w_d$ and $w_c$ smaller value weights for disc / cup) and background pixels ($w_b$ larger value weights for background) individually. This can reduce the difficulty of balancing the weights among multiple classes of pixels.  

\begin{equation}
MAE=\frac{1}{n}\sum_{j=1}^{n}\left | y_{j} - \hat{y}_{j}\right |
% MAE=\frac{1}{\sum w_j}\frac{1}{n}\sum_{j=1}^{n}w_j\left | y_{j} - \hat{y}_{j}\right |
\end{equation}
where n is the number of pixels; $\hat{y}_{j}$ is the predicted pixels; ${y}_{j}$ is the actual pixels.
% ;${w}_{j}$ is weights. 

% \subsubsection{Split-Copy-Merge}\label{split}
% Split-Copy-Merge belongs to the Divide and Conquer algorithm.  This strategy can make an earlier network prediction.  First,  we split the original label   into two image masks (1) Gray disc with white background (2) Black cup with white background.  Second, we train a X-Unet to predict the original label (gray-disc/ black-cup with white background), referred to X-Unet-O.   Third, we copy the learned weights of X-Unet-O to another two X-Unets  for predicting disc using the gray disc with white background training label (X-Unet-D) and cup using the black cup with white background training label  (X-Unet-C)  separately. Lastly, we merge the predictions of X-Unet-D and X-Unet-C to get the final segmentation mask result. 

\subsection{Classification}
The region and around-area of optic cup/disc contain the key pixel-wise features,  such as vertical disc diameter, the oval shape of disc/cup,  ISNT rule~\cite{harizman2006isnt}, and yellow-orange rim, that are mainly used for distinguishing glaucoma. The various scale pixel-wise features, including pixel color and location, are more important than high-level features, which can be learned by very deep convolutional neural networks (e.g., Resnet~\cite{he2015deep}).   Atrous (dilated) convolution~\cite{yu2015multi} is a key method that can extract different scale features and keep locations information simultaneously.  
\begin{figure}\label{deeplab}
\includegraphics[width=\textwidth]{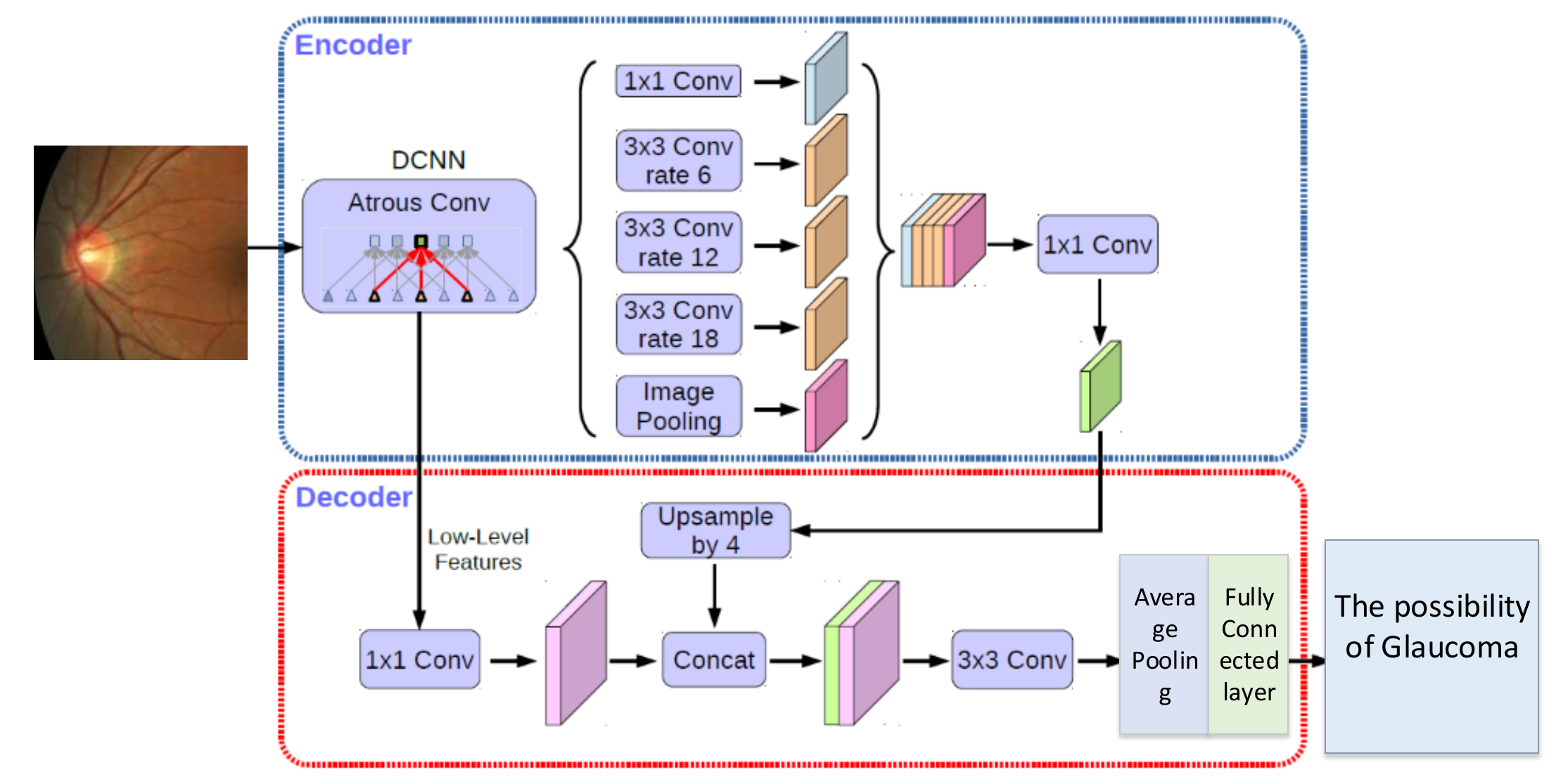}
\caption{Deeplab+3 variant architecture for glaucoma classification. We replace the last upsample layer with an average pooling layer and  a fully connected layer.}  
\end{figure}

We modify  DeepLab+3~\cite{chen2017rethinking} to be a classifier by replacing the last layer with a global average pooling layer followed by a fully connected layer for predicting the risk possibility of glaucoma.   DeepLab+3 includes one encoder and one decoder.  The encoder embeds atrous spatial pyramid pooling (ASPP) 
~\cite{chen2018deeplab} and convolutions in cascade to extract various scale context pixel-wise information.  The decoder refuses the low- level features learned by atrous convolutions with the various scale context features of the encoder. 

\subsubsection{Loss function}
We utilize binary entropy function to calculate the difference between the predicated class (possibility)  and actual class. 
\begin{equation}
BE=-(ylog(p)+(1-y)log(1-p))
\end{equation}
where BE presents the value of binary entropy loss; $y$ is the binary indicator (0 or 1); $p$ is the predicted probability. 

\section{Implementation details}

\subsubsection{Data prepossessing} 

We reduce the variance between training and validation images by cropping $600\times600$ size of region of interest (ROI) patches with the pre-trained model: Disc-aware Ensemble Network (DE-Net)~\cite{fu2018disc}. This data processing also can allow the model to focus on learning the most important pixel-wise information. 

\begin{figure}\label{crop}
\includegraphics[width=\textwidth]{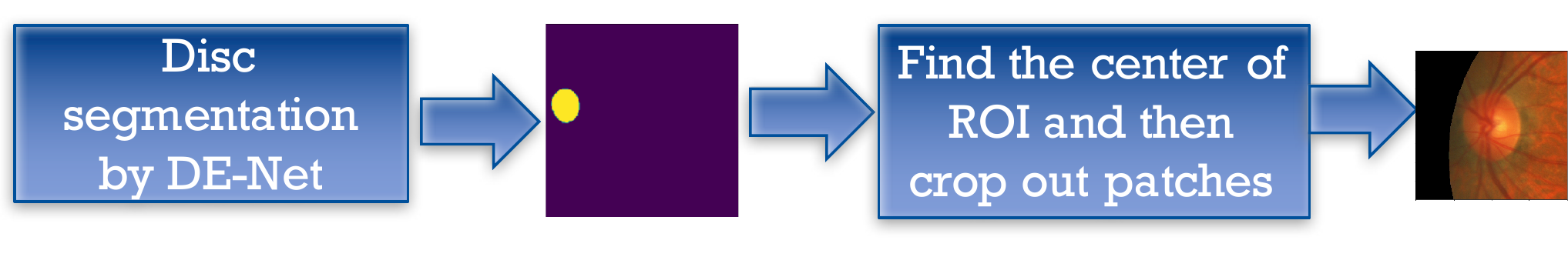}
\caption{The steps of data prepossessing for both segmentation and classification training and testing data.}  
\end{figure}

We use data augmentation skills, such as image Rotation -90/180/270  various angles and image flipping, to increase the number of training images.  In total, 3,200 ROI images are generated for segmentation and classification training.  

In order to ensure the network's receptive field is sufficient, we resize the training patches to be smaller size $128\times128$ as the segmentation task  training inputs. 
\begin{figure}\label{crop-classification}
\includegraphics[width=\textwidth]{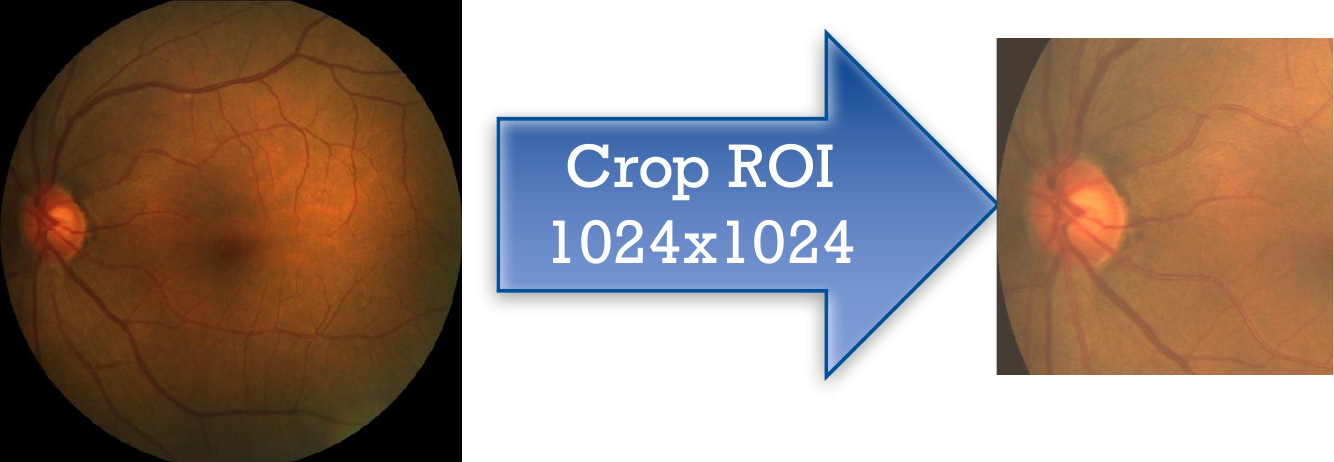}
\caption{The steps of data prepossessing for classification training data.}  
\end{figure}
For classification task, we use the same method (Fig.\ref{crop}) in segmentation to crop out the ROIs for training and testing. We resize the original cropped region images to be various sizes, such as $216\times216$, $256\times256$, $286\times286$, $324\times324$, and $360\times360$,  for multiple deep networks training. We average the models outputs as the final prediction result.

In addition, we need to handle with the image size difference between training and validation images. Hence, in testing stage on validation images, we crop out $500\times500$ (not the $600\times600$ in training) ROIs for segmentation task and $800\times800$ (not the $1024\times1024$ in training) ROIs for classification task. This can make sure the inputs to the network are similar to training images as much as possible. 

\subsubsection{Others} 
For training platform, we use Keras + tensorflow + python2.7.  The Adam optimizer is used and Learning rate is 0.0001.

\section{Results}
For segmentation,  on training set, mean Optic Cup Dice is 0.9626, mean Optic Disc Dice is 0.9876, and MAE CDR is 0.0161.
On validation set, mean Optic Cup Dice = 0.8498, mean Optic Disc Dice = 0.9433, and MAE CDR is 0.0444. 
Best rank (results-online): 8th.

For classification,  on training set  AUC: 1.0 and Sensitivity: 1.0.  Potential Overfitting is occured. 
On validation set AUC is  0.9708 and Sensitivity: 0.95. The latest (results-online) Rank: 2nd.  
\section{Conclusion}
In this work, we proposed two deep learning networks for retinal fundus glaucoma segmentation and detection. To overcome the major challenge, such as the variance (color and size due to different acquisition equipment) between training and testing images, we adopted pixel-wise learning and attention strategy, which can allow the networks focuses on learning the key features directly for the pixel-wise accurate predication. In particular, we proposed a multiple-input U-Net, named as X-Unet, to enlarge the raw image pixel information for low-level feature regression and prediction.  For classification, we proposed how to learn pixel-wise features for classification problems. In detailed, a dilated (Atrous) convolution based network can extract different scale features and keep locations information simultaneously. Atrous spatial pyramid pooling (ASPP) and convolution can extract various scale context pixel-wise information. The encoder part learns various pixel-wise features and the decoder part refuses the low- level features learned by dilated convolutions with the various scale context features of the encoder. 

Our proposed methods can overcome the variance issue between training and testing data.  However, we believe the best way to have a robust model is to standardize the image quality for any deep learning based model. This may need more efforts from both deep learning theorem and data acquisition community.

\small
\bibliographystyle{splncs04}

\bibliography{refer}
\end{document}